




\message{<< Assuming 8.5" x 11" paper >>}    

\magnification=\magstep1	          
\raggedbottom

\parskip=9pt

%

\def\singlespace{\baselineskip=12pt}      
\def\sesquispace{\baselineskip=16pt}      





%



\font\openface=msbm10 at10pt
 %

\def\Integers      {{\hbox{\openface Z}}}

 %
 %
 %



\font\german=eufm10 at 10pt

\def\Buchstabe#1{{\hbox{\german #1}}}









\def\supp{\mathop {\rm supp }\nolimits}

%

%
%



\def\implies{\Rightarrow}

%



\def\sqr#1#2{\vcenter{
  \hrule height.#2pt 
  \hbox{\vrule width.#2pt height#1pt 
        \kern#1pt 
        \vrule width.#2pt}
  \hrule height.#2pt}}




\def\lto{\mathop
        {\hbox{${\lower3.8pt\hbox{$<$}}\atop{\raise0.2pt\hbox{$\sim$}}$}}}
\def\gto{\mathop
        {\hbox{${\lower3.8pt\hbox{$>$}}\atop{\raise0.2pt\hbox{$\sim$}}$}}}
%
%
%


\def\half{{1 \over 2}}


\def\part{\subseteq}		

\def\braces#1{ \{ #1 \} }



\def\to{\mathop\rightarrow}	


\def\SetOf#1#2{\left\{ #1  \,|\, #2 \right\} }



\def\interior #1 {  \buildrel\circ\over  #1}     




\def\basisvector#1#2#3{
 \lower6pt\hbox{
  ${\buildrel{\displaystyle #1}\over{\scriptscriptstyle(#2)}}$}^#3}



\def\bar{\overline}		
 \let\miguu=\footnote
 \def\footnote#1#2{{$\,$\parindent=9pt\baselineskip=13pt%
 \miguu{#1}{#2\vskip -7truept}}}
%
%

\def\linebreak{\hfil\break}
\def\lbr{\linebreak}
\def\pagebreak{\vfil\break}


\def\BulletItem #1 {\item{$\bullet$}{#1 }}
\def\bulletitem #1 {\BulletItem{#1}}

\def\AbstractBegins
{
 \singlespace                                        
 \bigskip\leftskip=1.5truecm\rightskip=1.5truecm     
 \centerline{\bf Abstract}
 \smallskip
 \noindent	
 } 
\def\AbstractEnds
{
 \bigskip\leftskip=0truecm\rightskip=0truecm       
 }

\def\ReferencesBegin
{
 \singlespace					   
 \vskip 0.5truein
 \centerline           {\bf References}
 \par\nobreak
 \medskip
 \noindent
 \parindent=2pt
 \parskip=6pt			
 }
 %


\def\section #1 {\bigskip\noindent{\headingfont #1 }\par\nobreak\noindent}

\def\subsection #1 {\medskip\noindent{\subheadfont #1 }\par\nobreak\noindent}
%

\def\reference{\hangindent=1pc\hangafter=1} 

\def\ref{\reference}

 %

\def\journaldata#1#2#3#4{{\it #1\/}\phantom{--}{\bf #2$\,$:} $\!$#3 (#4)}
 %

\def\eprint#1{{\tt #1}}
 %
 %
 %

\def\webhome{{\tt http://www.physics.syr.edu/home1/sorkin/}}

\def\author#1 {\medskip\centerline{\it #1}\bigskip}

\def\address#1{\centerline{\it #1}\smallskip}

\def\furtheraddress#1{\centerline{\it and}\smallskip\centerline{\it #1}\smallskip}

\def\email#1{\smallskip\centerline{\it address for email: #1}}



\font\titlefont=cmb10 scaled\magstep2 

\font\headingfont=cmb10 at 12pt
%

\font\subheadfont=cmssi10 scaled\magstep1 
%







\def\A{\Buchstabe{A}}

\def\dom{\mathop{\rm dom}\nolimits} 

\def\Z2{\Integers_2}

\def\Star#1{#1^*}


\phantom{}



\sesquispace
\centerline{
   {\titlefont Quantum Dynamics without the Wave Function}\footnote%
   {$^{^{\displaystyle\star}}$}%
{ To appear in a special volume
  of {\it Journal of Physics A: Mathematical and General} entitled
 ``The Quantum Universe'' and dedicated to
 Giancarlo Ghirardi on the occasion of his 70th birthday.}}

\bigskip


\singlespace			        

\author{Rafael D. Sorkin}
\address
 {Perimeter Institute, 31 Caroline Street North, Waterloo ON, N2L 2Y5 Canada}
\furtheraddress
 {Department of Physics, Syracuse University, Syracuse, NY 13244-1130, U.S.A.}
\email{sorkin@physics.syr.edu}


\AbstractBegins                              
When suitably generalized and interpreted, 
the path-integral offers an alternative to 
the more familiar quantal formalism 
based on state-vectors, selfadjoint operators,
and external observers.
Mathematically one generalizes the path-integral-as-propagator
to a {\it quantal measure} $\mu$ 
on the space $\Omega$ of all ``conceivable worlds'', 
and this generalized measure expresses 
the dynamics or law of motion of the theory, 
much as 
Wiener measure expresses the dynamics of Brownian motion.
Within such ``histories-based'' schemes
new, and more ``realistic'' possibilities open up for 
resolving the philosophical problems of
the state-vector formalism.  
In particular, 
one can dispense with
the need for external agents
by
locating the predictive content of $\mu$ 
in its sets of measure zero: 
such sets are to be ``precluded''.
But unrestricted application of this rule 
engenders contradictions.
One possible response 
would remove the contradictions by circumscribing 
the application of the preclusion concept.
Another response, 
more in the tradition of ``quantum logic'',
would accommodate the contradictions 
by dualizing $\Omega$ to a space of ``co-events'' and 
effectively identifying reality with an element of this 
dual space.
\AbstractEnds


\sesquispace
\vskip -30pt

\section{} 
Reading the literature on ``quantum foundations'', you could easily get
the impression that the problems begin and end with non-relativistic
quantum mechanics.  
Perhaps 
most authors have 
limited themselves 
to this special case
because they
thought that no
essentially new 
philosophical questions 
arose in relativistic quantum field theory and
quantum gravity, or perhaps they merely felt that non-relativistic
quantum mechanics was hard enough to interpret without bringing in the
further complications of relative simultaneity and a dynamical causal
structure.  
This feeling
is true, no doubt, but it could also be
that an over-emphasis on the nonrelativistic case has led people in
directions they might 
not have taken 
had they thought
in a broader context.

Thus, for example, the concept of ``subsystem'' takes center stage in
many interpretations of the quantum formalism.  
In ``chemistry'' (the theory of nuclei, electrons, and
Coulombic interaction), one could perhaps construe a subsystem as a
definite collection of particles, but what could it mean in the context
of quantum field theory in curved spacetime, say?  Another example is
the concept of state-vector or ``wave function'', which seems to be
tied (in spirit if not always in the letter) to a particular moment of
time, while its unitary ``evolution'' takes place between designated
spacelike hypersurfaces.  
Locating one's basic mathematical structure on
a hypersurface is already uncomfortable in a relativistic spacetime, but
in relation to quantum gravity, it raises not only 
technical\footnote{$^\star$}
{The technical difficulties are especially severe in the context of
 causal set theory, where the obvious analog of a spacelike hypersurface
 is a maximal antichain.  But since an antichain has by definition no
 intrinsic structure of its own, one lacks an analog of the induced metric,
 which in the continuum is supposed to serve as the argument of the
 quantum gravitational wave function.  Extrinsically induced
 structures do exist and can be very useful for kinematic purposes
 as in [1];
 but they seem too ad hoc to enter into a basic dynamical law.}
but severe conceptual difficulties
(the notorious ``problems of time'').

For reasons such as these, anyone who thinks seriously about quantum
gravity, must at some point ask 
themself
whether the notion of
``history'' doesn't furnish a better starting point than the
Hilbert spaces, state-vectors and operator-observables that are so
familiar from textbooks.  
Or to pose the question another way: Is
quantum mechanics about the wave function and its unitary evolution or
is it about things like electrons, quarks and electromagnetic fields?
And if the latter, then is an electron to be described
three-dimensionally as (for example) a time-dependent point of space, or
four-dimensionally as a 
world line (not necessarily completed)?  
A ``realistic'' and  ``spacetime'' approach 
of the sort suggested by the last two sentences
may be termed ``histories-based''.  
Such an approach will of course  diverge technically from 
the state-vector approach,
but more importantly for this paper, 
I'm claiming
that it also differs at the purely philosophical,
interpretive level, and specifically that it suggests new
solutions to some of the recurring philosophical puzzles 
that have sprung from the quantum formalism.

To produce a histories-based formulation of quantum theory
means in essence to emancipate the path-integral from its role as an
auxiliary mathematical device 
(a mere tool for computing propagators or $S$-matrix elements) 
and to make of it the fundamental dynamical object of the theory.  
To that end, one must bring out more explicitly what it is that the path
integral really computes, and
having done so,
one must then complete the story by 
explaining how this newly conceived path integral 
can play the role played in classical mechanics by
the ``laws of motion'' ---
how in particular it can make possible the manifold 
(though still incomplete)
predictions that lend quantum mechanics its great practical utility.

Fortunately, much of this work of reformulation has already been carried
out; and we can claim to know --- or at least to have a very good
candidate for --- what this reconceived path-integral is in mathematical
terms.  
It is a mathematical object that can be represented either as a sort of
quadratic measure on $\Omega$ or as a Hermitian measure on
$\Omega\times\Omega$,
where $\Omega$ is the ``sample space'' in the sense of probability
theory.

\section{II.~ The path integral as a quadratic measure} 
%
Any theory for which a path integral can be defined possesses a
{\it sample space} $\Omega$ 
over which the integration takes place;\footnote{$^\dagger$}
{Fermionic path ``integrals'' are excepted here.  As normally defined, they
 are not integrals at all.  See however [2]}
but 
the kind of element $\gamma$ that comprises $\Omega$ varies from theory
to theory.  
In $n$-particle quantum mechanics, for example, each $\gamma$ would be
a set of $n$ trajectories; while for a scalar field theory, 
it might be a real or complex function  on spacetime.
Often the members of $\Omega$ are called ``fine-grained histories''
or simply ``histories'';
sometimes they are called ``paths'' or ``trajectories''.  
From the perspective of classical logic,
$\Omega$ is the space of ``possible worlds'', 
since each of its members represents
as complete a description of physical reality as is conceivable
in the theory.

Now the path integral is normally conceived of as a complex measure on
$\Omega$ (or more accurately, on subsets of $\Omega$ consisting of
trajectories with fixed endpoints).  As such, it serves to evolve
wave functions by giving the net amplitude for 
(say) a particle to
``propagate'' from one spacetime point to another.  But in quantum
mechanics as ordinarily understood, such amplitudes, and indeed the
wave function itself, are only intermediaries in the computation of
probabilities.  The object that yields these probabilities directly is
what I 
will call 
the quantal measure $\mu$, a function that assigns to
a subset $A\subseteq\Omega$ a real number $\mu(A)\ge0$.
Since this
characterization is a bit vague,  
let us examine it more closely in the context of
nonrelativistic point-particle mechanics, where the expression for $\mu$
is relatively simple.  

To that end, let $A$ be a subset of $\Omega$ defined by 
properties of $\gamma$ that refer only to 
its restriction to the time-interval $t\in[t_0,T]$; 
and let  $\rho(q,\bar{q})$ be the density-matrix at time $t_0$.  
Then the quantal measure of $A$ is given by\footnote{$^\flat$}
{When non-bosonic identical particles are involved, 
 one must insert into (1) an additional global phase-factor
 $\chi(\gamma,\bar\gamma)$ 
 that depends on the topological class to which the combined path
 $\gamma\cup\bar\gamma$ belongs.}
$$
  \mu(A) = \int\limits_{\gamma\in A^T} d\nu(\gamma)
           \int\limits_{\bar\gamma\in A^T} d\nu(\bar\gamma)
           \ 
       e^{iS(\gamma)-iS(\bar\gamma)}
          \ 
       \delta(\gamma(T),\bar\gamma(T)) 
          \ 
       \rho(\gamma(t_0), \bar\gamma(t_0))
       \ ,
       \eqno(1)
$$
where 
$A^T$ denotes the set of all truncated trajectories that can be derived from
  elements of $A$ by restriction to $[t_0,T]$,
$S(\gamma)$ is the action $\int Ldt$  of $\gamma$,
$\nu$ is some ``base measure'' on 
the space $\Omega^T$ of truncated histories
with respect to
which the integration is performed,
and
$\gamma(t)$ denotes the location (in configuration space $Q$) 
of $\gamma$ at time $t$.
(Notice that the ``truncation time'' $T$ can be chosen arbitrarily, as
long as it is late enough for it to be decided by then whether or not
$\gamma\in A$.)
It is then easy to check that, according to the standard quantum rules,
if the result of a measurement is
expressed by the position of a ``pointer'' 
at a given time,
and if $A_i$ is the set of histories for which 
the atoms composing the pointer are in the $i$th position
at that time,
then $\mu(A_i)$ is the
probability of that experimental outcome.
(Strictly speaking, 
 this interpretation of $\mu(A_i)$ as an objective probability
 goes beyond the standard rules,
 which only tell us that $\mu(A_i)$ is
 the probability that the pointer would be found in the $i$th position
 if measured by an observer external to the entire experimental setup.
 However, our only purpose here is to gain some familiarity with $\mu$
 and some intuition for its practical meaning.  
 Later we will attempt a more general and direct interpretation of $\mu$ 
 not based on the --- essentially classical --- concept of probability.)

Now the whole point of introducing the quantal measure was that it 
is histories-based and it
stands on
its own two feet, without needing to refer to the apparatus of operator
algebras, 
expectation-values,
state vectors, etc. that constitutes the terrain on which most
discussions of quantum philosophy take place.
We'd thus like to characterize $\mu$ in a general manner, without referring
to any specific expression like (1) above.
From a mathematical point of view, 
what makes quantum physics different from classical physics (in
which I'm including stochastic processes)
is its ``quadratic nature'': 
probabilities are squares of sums of
elementary amplitudes,
rather than simple sums of elementary probabilities. 
Physically, this corresponds to the phenomenon of {\it interference},
and the quadratic character simply expresses the fact that quantum
interference involves pairs of alternatives, but never triples (except
indirectly as induced by the pairwise interference).  
For the purposes of this article, the quadratic character is not 
absolutely essential, but as it expresses a very deep feature of quantum
mechanics, is seems appropriate to dwell on it for a bit.
This will also let me define a certain condition of ``strong
positivity'' that seems to play an important role in the detailed
working out 
of any interpretation of the quantal measure.

The quadratic nature of $\mu$ can be captured in two ways, either
directly or in terms of
a function $D$ which is
effectively a complex measure on 
$\Omega\times\Omega$.  
The former characterization 
[3]
rests on the
following sum rule 
for mutually disjoint subsets $A$, $B$, $C$ of
$\Omega$, which 
generalizes the simple additivity of classical probabilities:
$$
  \mu(A \sqcup B \sqcup C)
  - \mu (A \sqcup B) - \mu (B \sqcup C) - \mu (A \sqcup C)
  + \mu(A) + \mu(B) + \mu(C) = 0
  \ .
  \eqno(2)
$$
(The symbol $\sqcup$ denotes disjoint union.)
This sum-rule 
sits at level 2 of a hierarchy of 
sum-rules [3]
limiting
the ``degree'' of $\mu$.  
A classical probability-measure $\mu$ is
additive and 
resides at level 1;
a quantal $\mu$ resides at level 2,
which includes level 1 as a special case;
etc.\footnote{$^\star$}
{A function $\mu$ satisfying (2) is 
 quadratic in the following sense.  
 One can view
 $\mu$ as a kind of ``integral'' taking
 0-1-valued step functions to complex numbers; 
 and then (2) is
 the condition that this integral extend to a 
 homogeneous
 quadratic form on linear
 combinations of the step functions.
 (This does not contradict the fact that level 2 includes level 1.  If
 $\mu$ is additive it will of course also extend to a homogeneous {\it linear}
 form on linear combinations of step functions.)}


It bears emphasis that (2) is 
an unconditional
relationship that can only fail if quantum mechanics in the most general
sense fails.\footnote{$^\dagger$}
{Notice in this connection that whereas (1) produces complex
 amplitudes and is consistent only in the presence of unitarity,
 (2) presupposes neither of these features.  (Conversely, it
 sheds no light on {\it why} these features occur in nature.)}
It can be thought of concretely in terms of a three-slit diffraction 
experiment, and 
conversely one could perform such an experiment as a very
direct
null test of the validity of the basic quantal assumptions. 
More generally, 
any experiment in which three mutually exclusive
alternatives were made to interfere would provide a similar null test.
 
The above sum-rule expresses the quadratic nature of $\mu$ very
directly, but it is not as simple as one might wish.  
Fortunately, one can solve it 
identically
with the ansatz, 
$$
    \mu(A) = D(A,A)		\eqno(3)
$$
where $D$ 
(called for historical reasons a {\it decoherence functional}) 
will be required to fulfill the following axioms: 
\smallskip\noindent
{\it Hermiticity} \  $D(A,B) = D(B,A)^*$  \lbr
{\it Additivity} \   $D(A\sqcup B, C) = D(A,C) + D(B,C)$  \lbr
{\it Strong Positivity} \  
 for any finite collection of
subsets $A_1,A_2,\dots A_n$,
the $n \times n$
Hermitian
matrix $M_{ij} =D(A_i, A_j)$ is positive
semidefinite
(it has no negative expectation values).

To illustrate these axioms, 
let me just mention a couple of lemmas they imply,
which are useful in connection with the notion of preclusion that we
will turn to shortly:
\item{$(i)$} $\mu(A)=0 \ \implies\ D(A,B)=0$, whatever $B$ may be;
\item{$(ii)$} $\mu(A)=\mu(B)=0 \ \implies\ \mu(A\cup B)=\mu(A\cap B)$.

\noindent
(The first is a consequence of strong positivity; the second follows
from the first together with a certain identity that generalizes the
classical principle of ``inclusion-exclusion''.)

Having made the acquaintance of the quantal measure and its associated
decoherence functional, we now have the general form of a
histories-based quantum theory.  
The kinematics of such a
theory\footnote{$^\flat$}
{I almost wrote ``the ontology of such a theory'', but we will see 
 that this would have been too hasty.}
is expressed through its sample space $\Omega$; 
its dynamics is expressed through $\mu$ and $D$.
What is still lacking,
however,
is a 
finished
conceptual bridge to take us from this formal
structure to its 
physical meaning --- in particular to clarify how one can 
base predictions on $\mu$ without falling back on the problematic notions
of macroscopic system or external observer.

\section{III.~ The preclusion concept: toward an interpretation of $\mu$}
We have already introduced the sample space $\Omega$ and the quantal
measure $\mu$, 
which assigns non-negative real numbers to certain\footnote{$^\star$}
{Exactly which subsets of $\Omega$ belong to the domain of $\mu$ is a
 technical question whose answer will vary from theory to theory.
 In the special case where $\Omega$ is a finite set, 
 all of its subsets 
 would normally be included in the domain.  
 (Classically, the subsets in $\dom(\mu)$ are said to be ``measurable''.)}
subsets of $\Omega$.  
In the parlance of probability theory, such a subset is called
an {\it event}, and I will follow this usage,
although the phrase ``potential event'' might
be less subject to misinterpretation.
As subsets,
the events combine with each other in such a way as to
form a {\it Boolean algebra} $\A$,
and this algebraic structure will play a major role 
for us later in connection with  ``quantum logic''.

We have already seen that for some 
--- more or less well-defined --- 
set of ``instrument events'' $A$, \  $\mu(A)$ can be interpreted as an
experimental probability.  
This furnishes an important ``boundary
condition'' that must ultimately be satisfied by any interpretation that we
come up with, but could it be more than that?
Suppose
that one could identify within $\A$ a subalgebra $\A^{macro}$ consisting of
``the macroscopic events'', and that the restriction of $\mu$ to
$\A^{macro}$ could be proven to obey not only (2) but also the
level-1 sum rule, 
$\mu(A\sqcup B)-\mu(A)-\mu(B) = 0$.  
In that case, $\mu$ would define a probability functional on $\A^{macro}$
and one could try to maintain that this exhausted its physical meaning.
In doing so, 
one would be adopting the standpoint of ``decoherent'' or
``consistent'' histories [4].
But in this direction one 
encounters a series of problems, of which the first and most troublesome
at a practical level is the difficulty in coming up with a sufficiently
precise and tractable definition of ``macroscopic'' (that is of the
subalgebra $\A^{macro}$).  Lacking this, one risks obtaining either no
interpretation at all 
(if no satisfactory subalgebra can be shown to exist) or a
multitude of conflicting
interpretations (if more than one $\A^{macro}$ offers itself).
One might also worry 
(with a touch of whimsy)
that an interpretation of this sort 
denies a basic aspect of quantum mechanics because it
postulates,
essentially as a matter of definition, 
that alternative macroscopic
events cannot interfere, even in principle.  
More serious to my mind, 
though, 
is the fact that,
in itself,
such an interpretation
fails to offer us a picture of quantum reality
(this being arguably the most important 
task of any interpretation).
Or to the extent that it does suggest such a picture, 
that picture tends to deny the existence of the microworld.
It leads us to identify reality not with 
an individual history $\gamma$, but with an element of 
$\A^{macro}$, and this would withhold meaning from any 
statement referring to individual atoms or other forms of microscopic
matter. 

For reasons such as these, 
I do not think that we 
could rest content 
with any
interpretation founded on a distinction between $\A$ and $\A^{macro}$,
even if it could overcome the technical problems alluded to above.  
But
is there an alternative?
Can we draw predictive inferences from $\mu$ without invoking the word
``macroscopic''?   
To do so, 
we need to move away from the notion of probability, 
and one way of doing so is to replace it by
the notion of ``precluded event'', 
this being an event that ``to all intents and purposes cannot happen''. 
Indeed, 
probability is reducible to this notion
on one view
of the matter,
whence to take preclusion as the basis of our interpretation of the
quantal measure would only be
to 
carry over
to the quantal situation,
a time-honored idea for interpreting 
$\mu$
in the classical world.  
The basic idea, then, is to reduce all predictions to
statements of the form: 
``The event $A\in\A$ has zero measure $\mu(A)$, therefore $A$ is precluded''.  
Such a statement applies as well to microscopic events as to macroscopic ones.
The prediction which it makes is 
of a type that one might call ``definite but incomplete''.

\section{IV.~ The antinomy and three responses to it} 

\subsection {The problem with preclusion}

The principle we are aiming to hold onto affirms that events of measure
zero (or of sufficiently small measure) ``do not happen''.  
Implicit in any principle like this is some conception 
of reality,
with respect
to which one could say, in any given case, whether or not a given
$A\in\A$ did or did not actually happen;
and the most obvious and straightforward conception is the following.
Reality (that which exists or happens) 
would be a definite member $\gamma_{true}$ of
$\Omega$, and the statement that an event $A\subseteq\Omega$ 
``did not happen'' would simply mean that $\gamma_{true}$ was not to be
found among the elements of $A$.  
(If the event ``rain today'' is not happening,
that simply means that the ``actual world'' $\gamma_{true}$ is not among
those ``potential worlds'' $\gamma$ in which it is raining today.)
This particular ``logic of being'' 
is so ingrained in us that it is at
first hard to think of an alternative;\footnote{$^\dagger$}
{Fay Dowker has named as ``Axiom 0'' the assumption that
 one and only one member of $\Omega$ is realized.}
nevertheless it may
well 
turn out
that we need an alternative if we are to develop the preclusion
concept in a satisfactory manner.  The problem, in a nut shell, is that
there are ``too many preclusions''.  

An elementary illustration of the difficulty is furnished by the three-slit
experiment
referred to earlier, which we may idealize for present purposes
as a source emitting spinless particles which impinge on a diffraction
grating with three slits, labeled $a$, $b$ and $c$.
Let $A$ [respectively $B$, $C$]
be the set of $\gamma$ such that the particle traverses
slit $a$ [respectively $b$, $c$]
and arrives at
$p$, 
where $p$ is a spacetime
region---idealized as a point---which is aligned with the central slit and
which consequently sits within a ``bright band'' of the diffraction pattern.  
For such a $p$,
the measure
$\mu(A\cup B\cup C)$ 
of the set of all world lines that arrive at $p$ 
is nonzero,
corresponding to the fact that 
if we look for the particle at $p$,
we will frequently find it there.  
On the other hand, 
we can choose the separation between the slits so that, 
when taken in pairs $(a,b)$ or $(b,c)$, 
the amplitudes cancel, and correspondingly, 
the measures $\mu(A\cup B)$ and $\mu(B\cup C)$ will also vanish.
An unrestricted preclusion rule would then entail that the actual 
trajectory
$\gamma_{true}$
could belong neither to $A\cup B$ nor to $B\cup C$, 
whence it could not belong to $(A\cup B)\cup(B\cup C)=A\cup B\cup C$,
whence it could not arrive at $p$ at all---a false prediction.  
Although in this case, the particle could still go elsewhere, 
the Kochen-Specker setup gives rise 
to an example [5]
where 
every possibility without exception would be ruled out by an unrestricted
preclusion rule.
A theory cannot be more self-contradictory than that!

\subsection {First response: it's here, it's queer, get used to it!}

In a certain sense the same paradox is already present 
in classical probability theory, 
where for example, 
each single Brownian path $\gamma$ taken individually is of measure zero,
and therefore ``cannot occur'' ---
or ``almost surely'' cannot occur ,
in the stock phrase that mathematicians favor. 
In strict logic, 
this would seem to be a blatant inconsistency
since one $\gamma$ really does occur,
but it hasn't stopped people from applying the probability 
calculus to good effect.
They have just grown used to such paradoxes and learned to live with them. 

Could something similar happen with the paradoxes flowing from quantal
preclusion? 
Granted, they are made quantitatively worse by the effects of
interference (one needed an infinite number of preclusions to arrive at
a contradiction in the Brownian motion example, but only two in the
three-slit example), 
but could we nonetheless grow used 
to them and gradually cease to be troubled by them?\footnote{$^\flat$}
{Bob Griffiths seemed to recommend this as a long-term goal in a seminar
 at Perimeter Institute.}
If that were possible, we could embrace the unrestricted preclusion rule
with no worse a conscience than we feel in connection with the paradoxes
of ordinary probability theory.  Perhaps future generations will think
this way, but I personally would not bet on it.  
It might even turn out that, 
on the contrary, 
the effort to understand the quantal measure
will end up clarifying the meaning of classical probability, given that
the latter is but the classical shadow of the former.

\subsection {Second response: limited preclusion}

A second way out of the impasse might be to restrict the application of
the preclusion rule sufficiently so that no contradictions remained
among the surviving preclusions.  In another place [6], I've
written about this approach and its relationship to that of the
decoherent historians.  
I will not attempt to reproduce those reflexions
here, but only to append a few remarks on the problems and remaining
possibilities of the approach, 
which takes inspiration from the apparent absence of contradictions 
among the preclusions induced by measurements.
One seeks to abstract from measurement-situations some feature
that could make sense even in the microworld, but still 
serve as a seal of consistency wherever it is found.
Specifically, one sees in a measurement a special type of correlation, 
and one retains only those preclusions which express such correlations.\footnote{$^\star$}
{The word ``diremption'' has been used for a correlation of this sort.
 It denotes a ``splitting of the whole into two parts''.  Strictly
 speaking, however, the type of three-way correlation utilized in
 [6] would have to be called something like a ``triremption''.}

Not much effort has been devoted to developing this approach,
unfortunately, but two potential problems have been identified,
and I would like to sketch them for the benefit of the reader who is 
familiar with reference [6] or has thought along similar lines. 
The first problem, mentioned already in [6], concerns
``conditioning'' on the past, or on one's knowledge thereof.  
The question is whether one needs to introduce into the scheme a notion
of ``conditional preclusion'' or ``conditional measure'', corresponding 
to the  classical notion of conditional probability.
Such a notion would express in a certain sense the ``collapse of
the state-vector'', but it has never been clear whether it 
really is
needed, or whether,
on the contrary,
all valid predictions could be made without it, on
the basis of ``absolute'' (unconditional) preclusions.\footnote{$^\dagger$}
{If the practitioners of the Bohmian interpretation or of the
 ``many-worlds'' interpretation are correct in their understanding of
 ``branch formation'', then a separate notion of conditional preclusion
 would be superfluous.  However, I have never been able to convince
 myself that distinct ``branches of the wave function'' are articulated
 as sharply as they claim they are.}
If it turns out that it is needed, then that will represent an important
lacuna in the interpretation.

The second problem, that of ``specious diremptions'', is harder to
describe without referring to the details of the scheme elaborated in 
[6].   It springs from an apparent defect in the way that
correlations were characterized therein, and it has the effect that 
essentially any event of measure zero 
gets re-validated as a preclusion, since it
can be embedded in what counts as a correlation. 
Evidently, nothing would be gained by such a scheme in the way of
consistency.
I tend to believe that the problem is only a technical
one that could be fixed by a more
careful definition of correlation, but a ground for doubt is
that the task is reminiscent of that of distinguishing 
a mere correlation 
between two variables 
from 
a true cause-and-effect relationship between them.

A final problem, or at least an inconvenience, is the extra
work required to identify those sets of measure zero that correspond to
correlations of the required type.  This is reminiscent of the
difficulty of identifying decohering variables in the consistent
histories approach.  In the context of quantum gravity, where 
even the causal structure is dynamical, 
such difficulties tend to evolve into problems of principle.
Certainly, life would be simpler if one could simply affirm that all 
subsets of measure 0 were precluded.

\subsection {Third response: a modified logic}

Let us try to adhere to the original idea, according to which, without
exception, the vanishing of $\mu(A)$ entails the preclusion of $A$.  
We have seen how this rule engenders contradictions of which a simple
example was provided by three-slit diffraction.  
In that example, 
the slits were labeled $a,b,c$, 
and
the available preclusions apparently entailed the
false conclusion that the particle could never arrive at $p$.
Imagining an instance in which the particle actually does go 
to $p$,
we have in effect a contradiction between the preclusion
of $A\cup B$ and the preclusion of $B\cup C$.  Can we accommodate such
contradictions in our logic?

Before trying to answer this question, let me 
introduce a notation that makes explicit
a distinction 
which might seem to be empty, 
but 
which logicians have routinely drawn
between a proposition or statement and its ``truth-value''.  
The above subset $A$, for example, 
corresponds to the proposition 
``The particle passes through slit $a$ and arrives at $p$'',
but merely by referring to $A$, 
one does not necessarily assert 
that this proposition is true.  
To do that, 
one assigns the proposition a truth-value of `true'.
Let us agree to express this by writing $\phi(A)=1$, 
with $\phi(A)=0$ then signifying 
that the proposition is false. 
This distinction between a proposition and 
the assertion of the same proposition 
could strike one as pedantic at best,
but it seems less so 
when one speaks in terms of 
the corresponding ``predicates'' or ``properties''
(the property in question in our example 
being that of ``traversing the slit $A$'');
and it sounds still less forced
when one employs the grammar of question and answer:
To the subset $A$ then corresponds the question,
{\it Does the particle pass through slit $a$ (and arrive at $p$)?}, 
and to $\phi(A)$ corresponds the answer to the question,
either {\it yes}=1 or {\it no}=0,
as the case may be.


Expressed in terms of $\phi$, our 3-slit 
contradiction appears as follows:
$\phi(A\cup B)=0$ and $\phi(B\cup C)=0$, therefore $\phi(A\cup B\cup C)=0$.
Hence we could avoid the impasse if the first two equations could hold without
the third holding.   
In question and answer form,
this would amount to the following
(thinking of the slits as arranged in the order $(a,b,c)$).

``Does the particle avoid the left two slits? Yes.''

``Does the particle avoid the right two slits? Yes.''

``Does the particle avoid all three slits?  No.''

\noindent
Or if we assume that ``the particle avoids $a$ and $b$'' 
always has the same truth-value
as ``the particle goes through $c$'', then the first
two lines become:

``Does the particle go through $c$? Yes.''

``Does the particle go through $a$? Yes.'',

\noindent
which exhibits the contradiction still more clearly.

To embrace the contradiction is thus to accept that
the particle can possess apparently incompatible attributes.
But it is not altogether necessary to use this language.
Instead, one can shift the emphasis
and conceive of reality 
as represented,
not by ``a $\gamma$ with contradictory attributes'',
but by $\phi$ itself.
Whichever mode of expression one adopts,
a new logic is involved, 
one in which the truth-value of, 
for example, `$X$ and $Y$' 
does not follow by any universal rule
from the separate truth-values of `$X$' and of `$Y$'.

Classical logic demands that $\phi$ be a 
homomorphism
between the
event-Boolean algebra $\A$ and the two-element Boolean algebra $\Z2$
of ``truth values'',  meaning that `$X$ \& $Y$' is true if and
only if `$X$' and `$Y$' 
are both true, with similar ``truth tables''
applying to the other logical connectives, `or', `not', `xor', etcetera.
Because one can prove (at least when $\Omega$ is a finite set) that any such
mapping
can be equated to a unique element of $\Omega$,
identifying reality with a homomorphism 
$\phi:\A\to\Z2$ 
ends up being
nothing
more than a fancy way of identifying it with an element of $\Omega$.
That is, it yields nothing new.
But more general conceptions of reality arise if one modifies either
$\phi$ or the space (of ``truth values'') to which it maps.\footnote{$^\flat$}
{Logics that generalize $\Z2$ seem to be called ``modal''. 
 Such modifications have been proposed by many authors, notably in [7].  
 ``Quantum logic'' in the original sense of these words seems to have consisted
 primarily in replacing $\A$ by the lattice of subspaces of a hilbert
 space.  It did not seem to put forward any definite proposal for $\phi$,
 concerning itself more with probabilities than truth-values.
 The emphasis herein will fall neither on $\A$ nor on $\Z2$, but on the
 character of the mapping 
 between them.}
In the next section,  
we will consider a generalization which retains the same
two  truth-values but
relaxes the condition that $\phi$ be a homomorphism.
Of the fundamental logical connectives, 
$\phi$ will preserve only the one known as ``exclusive or''.

\section{V.~ An illustrative proposal: reality as a co-event} 
Many variations 
on
the simple idea proposed above are possible, but all
of them will 
describe reality by means of 
a particular function
$\phi_{true}:\A\to\Z2$ which generalizes the $\gamma_{true}$ of
classical logic in the manner explained above.  
In  choosing among the possible generalizations, 
it is useful to construe the event-algebra $\A$ as
literally an algebra in the strict mathematical sense of a vector space
over a field, in this case the finite field $\Z2$ containing only the two
elements $0$ and $1$.  
To make this work, however, one must identify the 
algebraic sum
$A+B$, not
with the set-theoretic union $A\cup B$, 
but with what is usually called ``symmetric
difference'', 
which differs from 
$A\cup B$ when
$A$ and $B$ overlap:
$$
   A + B = (A \cup B) \backslash (A \cap B)
   =
   \SetOf {\gamma\in\Omega} {\gamma\in A\cup B, \gamma\notin A\cap B}
   \ .
$$
With the 
algebraic product 
$AB$ taken to be the set-theoretic 
intersection $A\cap B$, 
all the axioms for an algebra are then satisfied.
For example, 
multiplication distributes over addition: $A(B+C)=AB+AC$,
and every event $A$ has an additive inverse $-A$ 
(namely $A$ itself).
Note that with these definitions $\A$
becomes a {\it unital} algebra, its unit $1$ being
the whole sample space $\Omega$,
because $\Omega A = \Omega\cap A = A$.
Set-theoretic complementation
is then effected just by adding 1:
$\Omega\backslash A=1+A$.
(In writing things this way, I am not distinguishing between $A$
considered as a subset of $\Omega$ and $A$ considered as an element of
the algebra $\A$.  This should not lead to any confusion, but 
it does take a little getting used to, especially if
one further equates $A$ to
the corresponding predicate. 
One then has three distinct sets of notation for essentially the same
set of operations: the set-theoretical notations like ``$\cap$'', the
algebraic notations like ``$+$'' and the logical notations like ``\&''!)

That 
$\phi$ classically 
is a unital algebra homomorphism
signifies that it 
then
preserves sum, product and unit; that is
$$
\eqalign{
  \phi(A+B)&=\phi(A)+\phi(B)	\cr
  \phi(A B)&=\phi(A) \phi(B)	\cr
    \phi(1)&=1
  }
$$
I will call the first condition linearity, and the second
multiplicativity. 
Which of these conditions shall we retain --- if any --- and which shall we modify?
If we were to drop all of them, the resulting framework would be
essentially vacuous.  If we were to keep them all, we'd be back to
classical logic.  We've already narrowed down the field of possibilities
by choosing to retain $\Z2$ as the ``codomain'' of $\phi$ (rather than
generalizing it to, say, a finite field of characteristic two).
With this choice, $\phi(A)$ can be only $1$ (true) or $0$ (false). 
If we want to preserve the symmetry between subsets and their
complements (that is, if we don't want to be forced to decide 
for each predicate, whether to regard it as ``positive'' or ``negative''
in nature), then it's natural to assume that $\phi(1+A)=1+\phi(A)$,
which says that a statement is true iff its negation is false.
But this is a special case of linearity, 
as long as $\phi$ is unital.\footnote{$^\star$}
{If $\phi$ is not unital, then necessarily $\phi(1)=0$.  In this case
 linearity implies that
 a statement is true iff its negation is {\it true}!  In either case,
 the truth of a statement is determined by that of
 its opposite, preserving the desired symmetry.}
It thus seems natural to make the following choice: retain
linearity and drop multiplicativity.
It also seems natural to retain the condition that $\phi$ be unital, but
we will lose nothing by leaving that choice open for now.

Let us agree to call a linear function $\phi:\A\to\Z2$ a {\it co-event},
and to qualify $\phi$ as {\it preclusive} 
if it assigns $0$ to every $A\part\Omega$ of $\mu$-measure zero.
Reality, then, is supposed to be a preclusive co-event.\footnote{$^\dagger$}
{A linear mapping from $\A$ to the space of scalars $\Z2$ defines an
 element of the vector space dual to $\A$.  Hence the nomenclature
 ``co-event''.
 In terms of logical operations,
 the linearity of $\phi$ means that the truth-value of 
 ($A$ XOR $B$) --- its truth  or falsehood --- still
 follows from the separate truth-values of $A$ and of $B$,
 even though 
 --- because we have dropped multiplicativity ---
 the truth-value of ($A$ AND $B$) does not.
 (Here `XOR' is exclusive or).}

Before considering what further conditions we might want our co-events
to satisfy, 
let us see how 
linearity and preclusivity
work out in the three-slit
example described earlier.  
Since the subsets $A$, $B$ and $C$ are mutually
disjoint, their unions can be written 
 as simple 
sums.  
Then, because $\phi$ has been assumed to be preclusive,
we have as before (but with our new notation):
$\mu(A+B)=0$ implies $\phi(A+B)=0$,
and
$\mu(B+C)=0$ implies $\phi(B+C)=0$.
The deductive train leading to a contradiction, now gets switched to a
new track, however, 
because $\phi(A+B)=0$ and $\phi(B+C)=0$ 
no longer 
combine to 
imply $\phi(A+B+C)=0$.  
Instead they imply only that
$\phi(A+C)=0$, 
as proven by the brief computation,
$\phi(A+C)=\phi(A+B+B+C)=\phi(A+B)+\phi(B+C)=0+0=0$,
where I've used that $B+B=0$ and $\phi(0)=0$ always.\footnote{$^\flat$}
{The earlier deduction
 implicitly used multiplicativity, which, 
 had we retained it, 
 would yield $\phi((A+B)(B+C))=0$, and thence $\phi(B)=0$ since
 $(A+B)(B+C)=AB+AC+BB+BC=0+0+B+0=B$.
 From this follows $\phi(A)=\phi(A+B)=0$, $\phi(C)=0$, etc.}
If we assume further that the particle really does arrive at $p$, then
we have additionally $\phi(A+B+C)=1$, 
which implies in the same way that
$\phi(A)=\phi(B)=\phi(C)=1$.  
This, then, is the unique preclusive co-event
$\phi$ for which the particle arrives at $p$.
In particular there {\it is} such a co-event, unlike with classical
reasoning, which drove us to conclude that $\phi(A+B+C)=0$.
In this way, we have avoided the false prediction.

In the 3-slit example, 
the relevant Boolean algebra $\A$ is
finite dimensional, which, over the finite field $\Z2$, implies that $\A$
itself is a finite set.  Whenever this happens,  
one may conveniently represent a co-event as a subset of
$\Omega$.  
In order
to see this, 
note first that\footnote{$^\star$}
{If the singleton set $\braces{\gamma}$ belongs to $\A$ 
 (as it normally will when $|\Omega|<\infty$) 
 then it is an {\it atom} of $\A$.  
 More generally, any atom
 $x\in\A$ defines a 
 unital co-event $\Star{x}$ by $\Star{x}(y)=\Theta(xy=x)$.
 (Like $\Star\gamma$, such an $\Star{x}$ is actually a homomorphism.)}
any history $\gamma\in\Omega$
defines a (unital) co-event $\Star{\gamma}$
through the equation $\Star{\gamma}(A)=\Theta(\gamma\in A)$,
where the notation $\Theta(\gamma\in A)$ denotes $1$ if
$\gamma\in A$ and $0$ if $\gamma\notin A$.
Notice further that we can suppose
without loss of generality that 
$\Omega$
itself is finite,
and  
then the $\Star{\gamma}$
furnish a canonical basis for $\A$
regarded as a vector space over $\Z2$.
This means that any coevent 
$\phi$
can be 
expanded in terms of the $\Star\gamma$, 
and we may 
therefore identify it with the
terms that occur nontrivially
in the expansion.
In this limited sense 
a co-event is
nothing but a subset of the sample space 
(a subset of odd cardinality if we require that $\phi$ be unital).  
In general, however, $\Omega$ will
be infinite, and the co-events will not be expressible so simply
(although finite subsets of  $\Omega$ will still yield co-events).
In the 
3-slit
example, our unique co-event was just
$\phi=\Star{A}+\Star{B}+\Star{C}$
(where the notation presupposes that we've limited $\A$ to the algebra
generated by $A$, $B$ and $C$.)


Our three-slit example illustrates how one may avoid 
(or perhaps I should say accommodate) 
the antinomies engendered by 
combining preclusion with classical logic, 
but it is rather a rudimentary example, and
it certainly does not prove that things will always work out so nicely.
Lacking a general theorem to that effect, we can only try the basic
ideas out on other examples of interest.  Let us therefore work out 
one example more, 
namely
the EPRB-like gedankenexperiment introduced by Lucien Hardy,
which 
illustrates
almost every known interpretive difficulty.

Before taking up this example, however, we must decide how to deal with
the fact that not every preclusive co-event is really suitable as a
potential reality,
 but only those which, in a sense still to be fully
determined, are 
``elementary''.
Perhaps the simplest way
to recognize the problem is to return to classical probability, but with
the measure $\mu$ regarded now as a special case of a quantal measure,
to which our general scheme can therefore be applied.  
Classically, 
one may encounter sets of measure zero, but never
interference,
and consequently every subset of 
a precluded subset 
is also
precluded.  From this, it is not hard to conclude  
(once again taking
$\Omega$ to be finite) that any 
$\gamma$ 
whose measure is itself nonzero
yields the
preclusive co-event
$\Star\gamma$,
and the most general preclusive co-event 
is 
an arbitrary sum of these.
On the other hand, 
we know that the ``right answer'' in this case (we are just doing
classical physics!) is that reality should be identified 
with a {\it single} $\gamma$, 
not 
with a
set of them.
The difficulty arises because we have abandoned
multiplicativity, 
consequently allowing arbitrary sums of
preclusive events to survive, since they clearly are also preclusive.
But this surely is not appropriate.
(For example the sum could include co-events that differed
macroscopically from each other.)  
What we seem to want to do is to reduce such sums to their ``elementary
parts'' and keep only the latter on our roster of potential realities.
That is, we want to allow multiple terms in the sum only when 
they are mandated 
by quantal interferences, 
not otherwise.
One way to do this is by 
{\it  ordering} the co-events appropriately and keeping only the minimal ones.

To that end, let me define the {\it support} of a co-event as the set of
$\gamma$ that 
comprise
it when we regard it as a subset of $\Omega$.
This definition works when $\Omega$ is finite, but not otherwise. 
A more general scheme
is suggested in the appendix.
Anyhow, with this definition, let us say that $\phi_1\prec\phi_2$ 
when 
$\supp(\phi_1)\part\supp(\phi_2)$;  
that is, we attribute to the
co-events 
the order induced by set-theoretic inclusion on the subsets
of $\Omega$ to which they correspond.  
(For example,
$\Star{A}+\Star{C}\prec\Star{A}+\Star{B}+\Star{C}$.)
We can then limit our potential realities 
to those preclusive co-events which are {\it minimal}
among all preclusive co-events.
In the classical situation we were just considering, 
this yields exactly
the right answer, that only the ``atomic'' coevents of the form
$\phi=\Star\gamma$ are retained.  
In other words, 
we have found a condition which 
is indistinguishable from multiplicativity in the classical case,
but which will allow the kind of more general co-event 
that we need for the
quantal case.  
(For example, our 3-slit co-event 
$\phi=\Star{A}+\Star{B}+\Star{C}$ 
is certainly  minimal, since no other
preclusive co-events exist at all.)

We are now prepared to take up the Hardy Gedankenexperiment [8]. 
In a histories setting it is convenient to conceive of this experiment
in terms of the trajectories of a correlated pair of spin-$\half$ silver
atoms passing through successive Stern-Gerlach beam-splitters
followed by recombiners.
If the atom on the left (say) passes first through a 
$\sigma_z$-analyzer, then a recombiner, and then 
a $\sigma_x$-analyzer,
it can follow any of four possible trajectories that we can label with a
pair of $\pm$ signs,
and similarly for the atom on the right.
In all, we have then a
sample space $\Omega$ of $4\times4=16$ members 
which we can label
$[++++]$,\dots,$[----]$,
taking the signs in the order:
$z$-left, $z$-right, $x$-left, $x$-right.
For example,
$[++--]$ is the trajectory such that both atoms traverse the ``upper''
beams in their respective $z$-analyzers, and subsequently the ``lower''
beams in their respective $x$-analyzers. 
(Compare the description in [9].)
We further imagine an idealized source that emits the atoms initially,
with equal amplitudes for each of the pairs
$(\sigma_z^{left},\sigma_z^{right})=(++),\ (+-),\ (-+)$,
but with zero amplitude for the pair $(--)$.
The decoherence functional $D$ and quantal-measure $\mu$ are then
determined as usual.
(See [9].)

The sixteen elements of $\Omega$ produce a Boolean algebra $\A$ of 
$2^{16}=65536$ events, of which a typical example, consisting of three
trajectories, is\footnote{$^\dagger$}
{For the sake of notational simplicity, 
 I will not distinguish between
 an element $\gamma$ of $\Omega$ and 
 the corresponding singleton subset $\braces\gamma$, 
 which is an element of $\A$ but not strictly speaking of $\Omega$.
 Thus, for example, (4) denotes the element of $\A$ given by
 the set whose members are the three trajectories,
 $[++--]$, $[+---]$ and $[-+--]$.
 (An object like $[+--+]$ is not realistically an element of the
 full physical sample space anyway,
 because
 it really comprises several trajectories
 and because 
 there is more in the world than just a pair of silver atoms.  
 In a formalization that took this into account, 
 our 65536 events would comprise a proper subalgebra
 of the full physical Boolean algebra;
 $[+--+]$ would be an atom of this subalgebra;
 and sample-spaces as such would hardly 
 need to be
 be mentioned.)}
$$
   [++--] + [+---] + [-+--]  \ .       \eqno(4)
$$
Now this particular event is important because it turns out to be
precluded classically but not quantally
(like the subset $A+B+C$ in our three-slit example).  
Unfortunately there is no space here to review why this fact is
interesting for ``hidden variables'' theories,
nor is there space even to do justice to the many interesting
mathematical and causal aspects of the gedankenexperiment as an example
of quantal measure theory.
Instead I will just enumerate the results and leave their further
analysis for another time.

By ``results'' I mean here the determination of the measure-zero events
and of the resulting minimal preclusive co-events.
The former 
(or at least all that I know of)
turn out to be the following:
$$
          [- - x y]   \ ,
$$
$$
            [+ + x -] + [+ - x -],  \quad [+ + - x] + [- + - x] \ ,
$$
where each `$x$' or `$y$' may be replaced by either `$+$' or `$-$'.
(In addition, as follows from the lemmas of \S II,
any $A$ which is a disjoint sum of these also satisfies 
$\mu(A)=0$, 
but 
by linearity 
that adds nothing to the determination of the preclusive
co-events.)
The preclusive co-events are 
those that map these eight
precluded events to $0\in\Z2$.
Among them are eight minimal ones, namely
$$
			     \Star{[++++]}
$$
$$
		\Star{[+-++]}    \qquad    \Star{[-+++]}
$$ 
$$
		\Star{[-++-]}    \qquad    \Star{[+--+]}
$$ 
$$
	     \Star{[++--]} + \Star{[+---]} + \Star{[-+--]}
$$
$$
  \Star{[+++-]} + \Star{[+-+-]}  \qquad \Star{[++-+]} + \Star{[-+-+]} 
$$
Of these,
the first five are ``classical'' 
(belonging to a single element of $\Omega$),
and the last two fail to be unital, since they contain an even number of
summands.
The sixth is thus the unique
minimal preclusive co-event 
which remains unital
while going beyond classical logic.
The most important thing to notice is that every ``observable event'' is
assigned `true' by some unital co-event in the list, 
where by an ``observable'' event I mean one that would be seen with
nonzero probability by an external agent who 
decided
to look for it.
In particular, the event ``both atoms in the lower beams after the
second splitting'' occurs
when
$\phi_{true}$ is 
the sixth co-event in the list.
In this sense our scheme has added to the classically possible co-events
only the bare minimum needed to be compatible with the experimental
facts, as predicted by
standard quantum mechanics.

\section{VI.~ Further work and final comments}
The thesis of this paper has been that a quantum theory should admit a
self-contained formulation based first of all on a space $\Omega$ of
``generalized trajectories'' or ``histories'', and secondly on a
``quantal measure'' $\mu$ defined on $\Omega$.
In particular, I argued that new interpretive possibilities arise in such
a framework which promise to provide a more satisfactory concept of
``quantal reality'' than offered by 
either ``the ensemble of instrument readings'' 
or ``the wave-function $\psi$'' (with or without ``spontaneous collapse''),
or by  ``a single element of $\Omega$''.
To illustrate this contention, 
I proposed 
to situate reality in 
a minimal preclusive co-event 
as defined in section V, 
which is a ``truth-function'' 
$\phi$ mapping
$\A$ to $\Z2$, where $\A$, the domain of $\mu$, 
is a Boolean algebra of ``events'' (subsets of $\Omega$),
and $\phi$ is linear when addition is interpreted as the so called
symmetric-difference of subsets 
(corresponding to the logical connective, exclusive-or).
The ``dynamical laws'' governing $\phi$ 
flowed from a
condition of ``preclusivity'', 
which 
required
that $\phi(A)=0$ for any 
$A\in\A$ such that $\mu(A)=0$.
To help clarify the meaning of the proposal, we worked out two examples,
involving 3-slit diffraction and EPRB-like correlations.

However two important questions have been addressed only in the
most implicit manner.  The first question is ``How does the past
condition the future?'',
which gets at the formal basis of prediction in the scheme;
the second is ``How does probability get into the story?''.
Although a full answer to either question could of course be given only
in the course of an extensive further working out of the basic ideas, 
sketches of the answers are already implicit in the basic assumptions of
the scheme, as set out above. 

The answer to the first question resembles the answer that the Bohmian
scheme would have to give: 
a past reality sets a ``boundary
condition'' for future realities
because the future must prolong the past.
The set of possible events codified in $\A$ 
grows with time in the
sense that 
$\A(t_1)\subset\A(t_2)$ when $t_1<t_2$.
Conversely any coevent $\phi_2$ for $\A(t_2)$
induces by restriction a coevent for $\A(t_1)$,
and 
this relationship entails
a condition that $\phi_2$ must satisfy, once $\phi_1$
is given.  It is this consistency condition that expresses the influence
of the past on the future.\footnote{$^\flat$}
{Subtle and important issues arise here, in connection with the 
 {\it minimality} conditions we have imposed on $\phi$; for minimality
 is in general not ``stable under time-development''.  
 One can find good examples of this in the second gedankenexperiment of
 section V.} 
More generally (in quantum gravity) ``the past'' might no longer
correspond  to a simple subalgebra of $\A$, 
but information about it
would still limit
the possible $\phi$'s 
in an analogous manner 
e.g. by specifying that 
certain ``stem-events'' [10] occurred.

The answer to the second question is, 
as indicated earlier,
that probability doesn't ``get in'' at all;
it's supposed to be there already in the form of the 
``preclusion rule'',
as applied to events $A$ of sufficiently small measure $\mu(A)$.
In an ``actual ensemble'' of repetitions of the same experiment, 
outcomes with the wrong frequencies will (we hope) be precluded.
(Notice that this is not meant to introduce some classical probability
measure on the space of co-events.  Such probabilities as are meaningful
are supposed to follow from the {\it quantal} measure $\mu$ on $\Omega$,
together the preclusion rule and the interpretation of reality as a
co-event.) 
It remains to be seen whether one can capture all
valid statistical predictions in this way.

Let me close with a list of
open questions,
some relating to the
new
notion of 
quantal
reality proposed above, and some relating to
``histories quantum mechanics'' more generally.

\BulletItem Will we be forced to accept co-events 
 $\phi$
 which assign `true' to two or more macroscopically distinct
 alternative events?
(this being perhaps the greatest danger for the 
scheme of \S V, 
assuming no 
simpler
inconsistencies show up).

\BulletItem
Will the ``mutability of minimality'' lead to trouble with causality?
(this being perhaps the second greatest danger.)

\BulletItem
How can one (or should one) incorporate fermionic (Grassmann) 
variables 
into a histories formulation
of quantum field theory?

\BulletItem
Can the interpretive scheme of section V shed any light on whether
field-particle duality has a home in histories-based quantum field
theory?  (Perhaps it opens up a narrow window for this.)

\BulletItem
Does the quantal random walk defined in [11] have, as conjectured, a
continuum limit that reproduces the quantum mechanics of the free
non-relativistic particle?  (This random walk appears to offer one of the few
fully defined examples of a genuinely non-unitary quantal process.)

\BulletItem
How does relativistic causality manifest itself in a histories
framework?  
Does there exist a suitable notion of ``quantum screening off'', as essayed in
[9]?

\BulletItem
Would anyone care to do the three-slit null test of quantum mechanics
described in section II above and in references [3] and [6]?


\bigskip
\noindent
I would like to thank Joe Henson and Fay Dowker for a series of email
exchanges that led to the proposal of section V, 
and Jim Hartle for clarifying questions about it.
This research was partly supported by NSF grant PHY-0404646.

\section{Appendix. Definition of the order $\,\prec\,$ in the general case}
Given two coevents $\phi_{1,2}:\A\to\Z2$, we defined $\phi_1\prec\phi_2$
in the body of the paper by $\supp{\phi_1}\part\supp{\phi_2}$.
Although an analogous criterion can be formulated for infinite Boolean
algebras, it doesn't seem to be what we want.
Instead the following appears to be the appropriate generalization from
finite to infinite Boolean algebras:
$$
   \phi_1 \prec \phi_2 \ {\rm iff} \ (\exists S\in \A)(\forall A\in \A)
   \ \phi_1(A) = \phi_2(SA)
   \ .
$$
It is not difficult to see that $\prec$ thus defined is reflexive,
transitive and acyclic, and that it agrees with the  definition in the
main text 
when $\A$ is finite. 

\noindent {\it Added note:}
Later work 
by Yousef Ghazi-Tabatabai
shows that 
the linearity condition of \S V 
is too restrictive to accommodate certain examples, 
the simplest of which involves 4-slit diffraction.

\ReferencesBegin                             

\ref [1]
Seth Major, David Rideout, Sumati Surya,        
``On Recovering continuum topology from a causal set'',
\eprint{gr-qc/0604124}

\ref [2] 
G.F. De Angelis, G. Jona-Lasinio and V. Sidoravicius,
``Berezin integrals and Poisson processes'',
\journaldata{J. Phys. A: Math. Gen}{31}{289-308}{1998},
\eprint{cond-mat/9709053}

\ref [3] 
Rafael D.~Sorkin,
``Quantum Mechanics as Quantum Measure Theory'',
   \journaldata{Mod. Phys. Lett.~A}{9 {\rm (No.~33)}}{3119-3127}{1994},
   \eprint{gr-qc/9401003},\lbr 
   \webhome{};
\lbr
Roberto B.~Salgado, ``Some Identities for the Quantum Measure and its Generalizations'',
 \journaldata{Mod. Phys. Lett.}{A17}{711-728}{2002}
 \eprint{gr-qc/9903015}

\ref [4] 
J.B.~Hartle, ``Spacetime Quantum Mechanics and the Quantum 
 Mechanics of Spacetime'',
 in B.~Julia and J.~Zinn-Justin (eds.),
 {\it Gravitation et Quantifications: 
      Les Houches Summer School, session LVII, 1992}
 (Elsevier Science B.V. 1995)
 \eprint{gr-qc/9304006}
R.~B. Griffiths, 
 ``Consistent histories and the interpretation of quantum mechanics'',
 \journaldata{J.~Statist. Phys.}{36}{219-272}{1984}

\ref [5] 
Fay Dowker and Yousef Ghazi-Tabatabai (in preparation)

\ref [6] 
  Rafael D.~Sorkin,
``Quantum Measure Theory and its Interpretation'', 
  in
   {\it Quantum Classical Correspondence:  Proceedings of the $4^{\rm th}$ 
    Drexel Symposium on Quantum Nonintegrability},
     held Philadelphia, September 8-11, 1994,
    edited by D.H.~Feng and B-L~Hu, 
    pages 229--251
    (International Press, Cambridge Mass. 1997),
    \eprint{gr-qc/9507057},
    \webhome{}

\ref [7] 
C.J.~Isham, ``A Topos Perspective on State-Vector Reduction'',
\eprint{quant-ph/0508225}, and references therein.

\ref [8] 
Lucien Hardy, 
``Quantum mechanics, local realistic theories, and Lorentz-invariant 
 realistic theories'',
 \journaldata{Phys. Rev. Lett.}{68}{2981-2984}{1992};
and
``Nonlocality for two particles without inequalities for almost all entangled
 states'',
 \journaldata{Phys. Rev. Lett.}{71}{1665-1668}{1993}.

\ref [9]  
David Craig, Fay Dowker, Joe Henson, Seth Major, David Rideout and Rafael D.~Sork\-in,
``A Bell Inequality Analog in Quantum Measure Theory'',
\eprint{quant-ph/0605008},\lbr
\webhome{}

\ref [10] 
Graham Brightwell, {H. Fay Dowker}, {Raquel S. Garc{\'\i}a}, {Joe Henson} 
 and {Rafael D.~Sork\-in},
``General Covariance and the `Problem of Time' in a Discrete Cosmology'',
 in K.G.~Bowden, Ed., 	
 {\it Correlations}, 
 Proceedings of the ANPA 23 conference,
 held August 16-21, 2001, Cambridge, England 
 (Alternative Natural Philosophy Association, London, 2002), pp 1-17
\eprint{gr-qc/0202097},
\webhome{};
and 
Graham Brightwell, Fay Dowker, Raquel S.~Garc{\'\i}a, Joe Henson and 
     Rafael D.~Sork\-in,
``{$\,$}`Observ\-ables' in Causal Set Cosmology'',
\journaldata {Phys. Rev.~D} {67} {084031} {2003}
\eprint{gr-qc/0210061},
\webhome{}.
(These papers do not employ the term ``event''; instead they
refer to ``stem sets'', ``stem predicates'' or ``stem
questions''.)

\ref [11]
Xavier Martin, Denjoe O'Connor and Rafael D.~Sorkin,
``The Random Walk in Generalized Quantum Theory''
\journaldata {Phys. Rev.~D} {71} {024029} {2005}
\eprint{gr-qc/0403085},
\lbr
\webhome{}

\end               


(prog1    'now-outlining
  (Outline 
     "\f......"
      "
      "
      "
   ;; "\\\\message"
   "\\\\Abstrac"
   "\\\\section"
   "\\\\subsectio"
   "\\\\appendi"
   "\\\\Referen"
   "\\\\ref....[^|]"
  ;"\\\\ref....."
   "\\\\end